\documentclass[aps, reprint, prb, longbibliography]{revtex4-1}

\usepackage{graphicx}
\usepackage{amsmath}
\usepackage{wasysym}
\usepackage{amsfonts}
\usepackage{color}
\usepackage{verbatim}
\usepackage{hyperref}

\usepackage[final]{changes}
\graphicspath{{./figures/}}

\newcommand{\beq}{\begin{equation}}
\newcommand{\eeq}{\end{equation}}
\newcommand{\bea}{\begin{eqnarray}}
\newcommand{\eea}{\end{eqnarray}}
\newcommand{\bal}{\begin{align}}
\newcommand{\eal}{\end{align}}

\newcommand{\fig}[1]{Fig.~\ref{#1}}

\newcommand{\SiN}{Si$_3$N$_4$}
\newcommand{\Ag}{A$_{1g}$}

\begin{document}

\title{Coherent and Incoherent Structural Dynamics in Laser-Excited Antimony}

\author{Lutz Waldecker}
\email{waldecker@stanford.edu}
\altaffiliation[Current affiliation: ]{Stanford University,  348 Via Pueblo Mall, Stanford, California 94305, USA.}
\affiliation{Fritz-Haber-Institut der Max-Planck-Gesellschaft, Faradayweg 4-6, 14195 Berlin, Germany}
\author{Tobias Zier}
\affiliation{Theoretische Physik, Universit\"at Kassel and Center for Interdisciplinary Nanostructure Science and Technology (CINSaT), Heinrich-Plett-Stra{\ss}e 40, 34132 Kassel, Germany}
\author{Thomas Vasileiadis}
\affiliation{Fritz-Haber-Institut der Max-Planck-Gesellschaft, Faradayweg 4-6, 14195 Berlin, Germany}
\author{Roman Bertoni}
\altaffiliation[Current affiliation: ]{Universit\'e de Rennes 1, Institut de Physique de Rennes, 263 av. G\'en\'ral Leclerc 35042 Rennes cedex, France}
\affiliation{Fritz-Haber-Institut der Max-Planck-Gesellschaft, Faradayweg 4-6, 14195 Berlin, Germany}
\author{Felipe Valencia H.}
\altaffiliation[Current affiliation: ]{Physics Department, Universidad Nacional de Colombia, Edificio 404, Ciudad Universitaria, Bogot\'a, Colombia}
\affiliation{Theoretische Physik, Universit\"at Kassel and Center for Interdisciplinary Nanostructure Science and Technology (CINSaT), Heinrich-Plett-Stra{\ss}e 40, 34132 Kassel, Germany}
\author{Martin E. Garcia}
\affiliation{Theoretische Physik, Universit\"at Kassel and Center for Interdisciplinary Nanostructure Science and Technology (CINSaT), Heinrich-Plett-Stra{\ss}e 40, 34132 Kassel, Germany}
\author{Eeuwe S. Zijlstra}
\affiliation{Theoretische Physik, Universit\"at Kassel and Center for Interdisciplinary Nanostructure Science and Technology (CINSaT), Heinrich-Plett-Stra{\ss}e 40, 34132 Kassel, Germany}\author{Ralph Ernstorfer}
\email{ernstorfer@fhi-berlin.mpg.de}
\affiliation{Fritz-Haber-Institut der Max-Planck-Gesellschaft, Faradayweg 4-6, 14195 Berlin, Germany}

\begin{abstract}

We investigate the excitation of phonons in photoexcited antimony and demonstrate that the entire electron-lattice interactions, in particular coherent and incoherent electron-phonon coupling, can be probed simultaneously. Using femtosecond electron diffraction (FED) with high temporal resolution, we observe the coherent excitation of the fully symmetric \Ag\  optical phonon mode via the shift of the minimum of the atomic potential energy surface. 
Ab initio molecular dynamics simulations on laser excited potential energy surfaces are performed to quantify the change in lattice potential and the associated real-space amplitude of the coherent atomic oscillations. Good agreement is obtained between the parameter-free calculations and the experiment.
In addition, our experimental configuration allows observing the energy transfer from electrons to phonons via incoherent electron-lattice scattering events. The electron-phonon coupling is determined as a function of electronic temperature from our DFT calculations and the data by applying different models for the energy-transfer.

\end{abstract}

\maketitle
\date{\today}

\section{Introduction}

The excitation of electrons in molecules and solids can change their real-space electronic distribution and therefore the potential energy landscape of the nuclei. Examples include semiconductors, where photoexcitation drives transitions from bonding to antibonding states \cite{Rousse2001,Harb2008}, some metal oxides \cite{Bothschafter2013}, materials with ionic bonding \cite{Stingl2012}, charge density wave systems \cite{Eichberger2010} and Peierls-like distorted materials such as bismuth and antimony \cite{Zeiger1992}. A sudden change of the potential energy surface, as induced by photoexcitation of electrons with femtosecond laser pulses, induces collective atomic motion towards its new minimum, resulting in coherent lattice oscillations \cite{Silvestri1985} or even non-thermal phase transitions \cite{Sciaini2009}. The non-equilibrium state with electrons highly excited in an initially unperturbed lattice relaxes by incoherent transfer of energy from electrons to lattice vibrations until a new thermal state is reached. The rate at which energy is transferred to and between phonon modes depends on electron-phonon as well as phonon-phonon scattering processes \cite{Waldecker2016} and determines the lifetime of the non-thermal state. Both processes are therefore intimately connected and need to be investigated to understand the complex lattice dynamics following photoexcitation.

The crystal structure of antimony can be seen as a distorted cubic structure, in which the body diagonal is elongated. Along this direction, a Peierls-like distortion leads to a pairing of atoms and a doubling of the unit cell.  The widely accepted picture is that the electronic excitation, induced e.g.~by a femtosecond laser pulse, changes the potential energy surface along the Peierls-like distorted crystal-direction (z-direction). 
If the laser pulses are shorter than an oscillation period, the shift is quasi-instantaneous and the atoms start moving collectively. The mechanism is thus referred to as Displacive Excitation of Coherent Phonons (DECP). The excitation and the decay of these coherent lattice oscillations in antimony and bismuth has been observed in early optical pump-probe experiments \cite{Cheng1990} and has been studied extensively since then with the applied techniques ranging from all-optical measurements at various excitation conditions \cite{Hase1998, Stevens2002, Ishioka2006, Ishioka2008}, to time-resolved x-ray scattering \cite{KST2003, Johnson2008} and molecular dynamics (MD) simulations \cite{Zijlstra2006PROC}. 
The incoherent electron-phonon coupling in bismuth has been studied by density functional theory (DFT) \cite{Arnaud2013} and time-resolved electron diffraction \cite{KST2015}. Electron diffraction has also been applied to investigate the expansion \cite{Esmail2011} and acoustic breathing modes of bismuth \cite{Bugayev2011, Moriena2012}.

Where optical measurements offer a high temporal resolution and are comparably easy to perform, they only provide an indirect measurement of the lattice dynamics. 
Such experiments are sensitive to the dielectric function, which depends on the electronic structure and therefore indirectly on the state of the lattice. 
The disentanglement of electronic and structural dynamics from time-resolved optical measurements is not trivial and, specifically, the incoherent transfer of energy from electrons to the lattice can only be retrieved if the measurements are accompanied by theory calculations \cite{Katsuki2013}. 
Time-resolved x-ray diffraction experiments provide a more direct access to the lattice dynamics. Due to the small Ewald-sphere of hard x-rays in the 10's of keV range, the probed area in reciprocal space is small, allowing for the simultaneous observation of few Bragg peaks only. 
In contrast, the typical energies of electrons in femtosecond electron diffraction experiments of around 100 keV allow to map a large part of reciprocal space and thus a complete determination of the lattice dynamics. 
So far, however, the limited time-resolution of these experiments has only allowed investigating coherent and incoherent electron-phonon coupling for low-frequency shear or breathing modes~\cite{Harb2009,Chatelain2014}.

Here, we present femtosecond electron diffraction experiments on textured samples with sufficient time-resolution to simultaneously determine the excitation of coherent optical phonons and the incoherent energy-transfer from electrons to lattice vibrations in antimony.  
Complementary ab initio MD simulations are performed to connect the measured diffraction intensities with real-space atomic displacements. 

\section{Experimental Techniques}

\begin{figure}[bth!]
\centering
% reprint
\includegraphics[width=1.0\columnwidth]{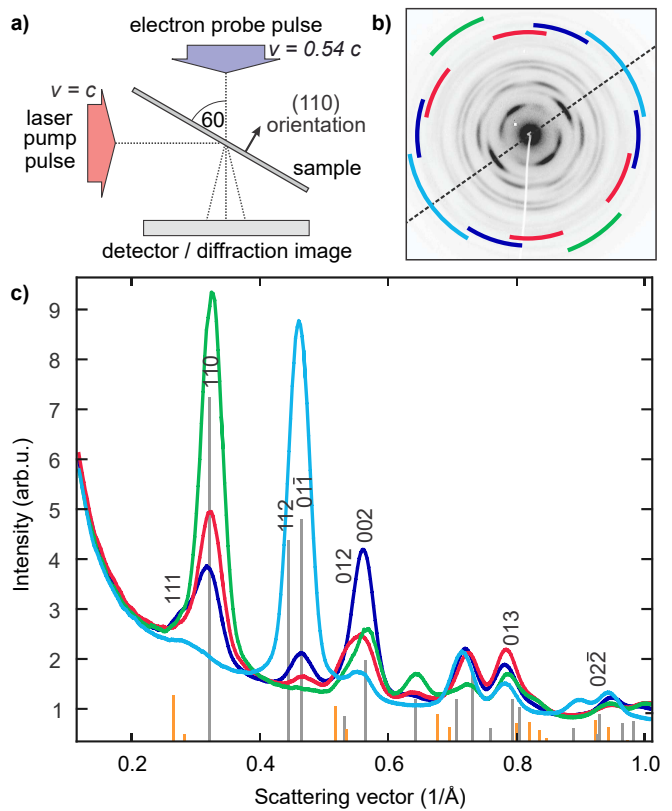}
\caption{{\textbf a)} Sketch of the experimental configuration. {\textbf b)} Diffraction image of tilted, polycrystalline antimony with (110) texture. The axis of rotation is shown as a dotted line. {\textbf c)}  Radial averages, calculated by integrating selected angular regions (colored bars) of the diffraction pattern shown in b). The different ratios of the intensity of closely spaced peaks allows a better determination of relative intensity changes with pump-probe delay (not shown). The lines indicate the positions of Bragg-peaks that remain (gray) or vanish (orange) in the more symmetric phase. }
\label{fig:raw}
\end{figure}

In the FED experiments, laser pulses of a 1 kHz amplified Ti:Sapphire laser system with 800 nm central wavelength and duration of 50 fs FWHM were used to excite free-standing samples. 
Diffraction images were taken with electron pulses arriving at the sample at variable delays with respect to the pump pulses. 
The electron pulses contain few thousand electrons and are accelerated to an energy of 94 keV, resulting in pulse durations of around 100 fs on the sample position~\cite{Waldecker2015Setup}. 
The samples were tilted by 30 degrees with respect to the electron propagation direction and the pump-pulses were directed onto the sample under an angle of 60 degrees, perpendicular to the electron propagation direction, see~Fig.\ref{fig:raw} a.
In this geometry, the mismatch in arrival time at the sample plane of probe electrons, which are moving with a speed of $0.54\cdot c$, and the optical pump pulse is minimized.

%%%%%%%%%%%%%%%% Figure 1 %%%%%%%%%%%%%%%%%%%%%%%%%%%%%%%%%%%%%
%%%%%%%%%%%%%%%%%%%%%%%%%%%%%%%%%%%%%%%%%%%%%%%%%%%%%%%%%%%

The samples used in this study were 30 nm thick films of Sb, sandwiched between two 5 nm thick films of \SiN. They were produced by successive sputtering of the different films onto a NaCl crystal. By dissolving the NaCl in water, the samples detach and float on the water surface, from where they were picked up with standard TEM grids. 

The deposition of Sb on a \SiN\ film resulted in polycrystalline samples with the crystallites showing a preferred orientation of their (110) axis parallel to the film normal, as evidenced by the vanishing intensity of the respective lattice peak in diffraction images taken at normal incidence of the electrons and in agreement with earlier reports \cite{Patel1979}. 
As the samples were tilted by 30 degrees compared to the electron propagation direction, their texture results in changes of the intensity of the polycrystalline diffraction rings in angular direction, as seen in the raw diffraction image in \fig{fig:raw} b. 
This allows identifying and separating closely spaced diffraction rings. Radial averages (RAs) were calculated from the raw images by integrating them in angular direction within four distinct areas, chosen to well separate some closely lying peaks. \fig{fig:raw} c shows the resulting diffraction patterns together with literature values of the Bragg-peak positions \cite{Barrett1963}. 
In the case of the Peierls-distortion being lifted, i.e.~the atomic positions being shifted to $z=0.25$ and $z=0.75$ along the c-axis of the unit cell, the symmetry changes, leading to the disappearance of some Bragg-peaks. Those peaks are indicated by orange lines in \fig{fig:raw} c and are referred to as Peierls-distortion (PD) peaks, whereas those that persist are referred to as high-symmetry (HS) peaks. 
As the \Ag\ phonon mode corresponds to an atomic motion towards (and away from) the more symmetric phase, the relative changes of the intensity of the PD peaks are particularly large.

The evolution of the Bragg peak intensities is obtained by fitting the radial averages taken at different pump-probe delays. 
First, the background, composed of inelastic scattering of Sb as well as scattered intensity of the \SiN\ films, is determined and subtracted. 
Second, the peaks are fitted with pseudo-Voigt line profiles. 
The evolution of the intensity of each Bragg peak is obtained from the RA in which it is most dominantly present. 

\section{Computational Methods}
\label{comp-meths}

For the theoretical simulation of the structural response of antimony to an ultrashort laser excitation we used the Code for Highly excIted Valence Electron Systems (CHIVES), which has successfully explained several ultrafast phenomena \cite{Zijlstra2013PRX,Zijlstra2013Adma,Zijlstra2014APA,Zier2015,Zier2016}. CHIVES is an ab initio code that uses the temperature dependent density functional theory formalism \cite{Mermin1965}, in which the electronic temperature $T_e$ of the system is the key quantity for describing the laser excitation. 
$T_e$ is set to 2000~K, which we estimate the initial electronic temperature to be in the experiment from the lattice temperatures reached at long time-delays and the electronic and lattice heat capacities (see Sec. \ref{results}). 
The electronic temperature is then kept constant during the $1.25$ ps of our MD simulation runs in the laser-excited state. 
Incoherent electron-phonon coupling reducing the electronic temperature is not considered in the MD simulations, which are therefore only valid for times at which the increase of ionic temperature through incoherent heating is small and does not much effect the coherent motion, i.e., $t\lesssim 1$ ps, see Fig.\ \ref{fig:TTM} c. 
On the other hand, the coherent electron-phonon coupling is fully taken into account by the change in the potential energy surface, computed by CHIVES on the fly at every time step. The exchange and the correlation energy functionals are computed in the local density approximation. For the electronic system, CHIVES uses norm-conserving pseudopotentials \cite{Hartwigsen1998} for the tightly bound core electrons and atom-centered Gaussian basis sets for the less bound valence electrons. The basis set for antimony consists of $18$ basis states \cite{Zijlstra2016}. 
Our supercell contains $N=1440$ atoms, which shows a good convergence in the quantities of interest and enables us to reach a high $q$ point resolution for the structure function, see also \cite{Zier2015}. 
We stress that these are the first ab initio MD simulations on laser excited coherent phonons in antimony using the full laser excited interatomic potential. For statistical reasons, we averaged over $20$ independent runs, where we initialized each run near room temperature using an Anderson thermostat.  

The comparison of the MD simulations to the experiment is done by calculating the intensities of different diffraction peaks directly from the atomic positions at various time steps of the simulation using  
\beq
\label{eq:sf}
I (t) = \left| \sum^{N}_{j=1} f_j e^{- 2 i \pi (h\mathbf{b_1} +k\mathbf{b_2}+l\mathbf{b_3})\cdot
  \mathbf{r_j (t)}} \right|^2 \; ,
\eeq
where $N$ is the number of atoms in the used supercell, $hkl$ are the Miller indices, $\mathbf{b_1}$, $\mathbf{b_2}$, $\mathbf{b_3}$ are the reciprocal basis vectors of the unit cell, and $\mathbf{r_j (t)}$ are the positions of the atoms at each timestep $t$.
In order to compute the electronic heat capacity, we calculated the change of the total energy by additionally performed static calculations at different electronic temperatures between $T_e = 280$ and $2650$ K. 

The calculations for the electron-phonon coupling follow the  Allen formalism (for details see ref \citenum{Waldecker2016} ), in which the coupling strength is approximated with the Eliashberg spectral function ($\alpha^2F$):
\begin{align}
\label{eq:gepcalc}
G_{ep}(T_e,T_l) = & -\frac{2\pi N_c\hbar^2}{T_e-T_l} \nonumber \\
&{\displaystyle\int\limits_0^\infty }d\omega \omega^2\alpha^2F(\omega)\left[n_B(\omega,T_e)-n_B(\omega,T_l)\right] \nonumber  \nonumber \\
              & \times{\displaystyle \int\limits_{-\infty}^\infty}d\epsilon \frac{g^2(\epsilon)}{g(\epsilon_F)}\frac{\partial f(\epsilon,T_e)}{\partial \epsilon},
\end{align}
where $n_B(\omega,T)$ and $f(\epsilon,T)$ are the Bose-Einstein and Fermi-Dirac distributions, respectively, and $g(\epsilon)$ is the electronic density of states. The spectral function and density of states $g(\omega)$ were obtained  with the Quantum-ESPRESSO package \cite{QE2009} using dense  Monkhorst-Pack grids of 28x28x28 {\bf k}-points and 14x14x14 {\bf q}-points in the rhombohedral unit cell. The differences between the values of the electron-phonon coupling $G_{ep}$ calculated using this fine sampling and sparser meshes of 14x14x14 {\bf k-} and 7x7x7 {\bf q-} points were only 6\%.
Here, the electronic system is described by norm conserving pseudopotentials but instead of Gaussian basis sets a plane wave basis is used with a cutoff energy of 60~Ry (816~eV). 
For these calculations we used the PBE approximation to the exchange correlation functional, and relaxed the lattice parameters and internal coordinates to their optimal values. 
We note, that the predicted cell volume at this level of theory is about 7\% larger than the experimental one and the internal coordinates are off by less than 0.2\%. The electronic and vibrational densities of states and, correspondingly, the electron and lattice contributions to the specific heats calculated with the PBE and LDA methods  are in good agreement with each other. We did not consider the spin-orbit interaction in our calculations. 

\section{Results}
\label{results}

%%%%%%%%%%%%%%%% Figure 2 %%%%%%%%%%%%%%%%%%%%%%%%%%%%%%%%%%%%%
\begin{figure*}[bth!]
\centering
\includegraphics[width=1.0\textwidth]{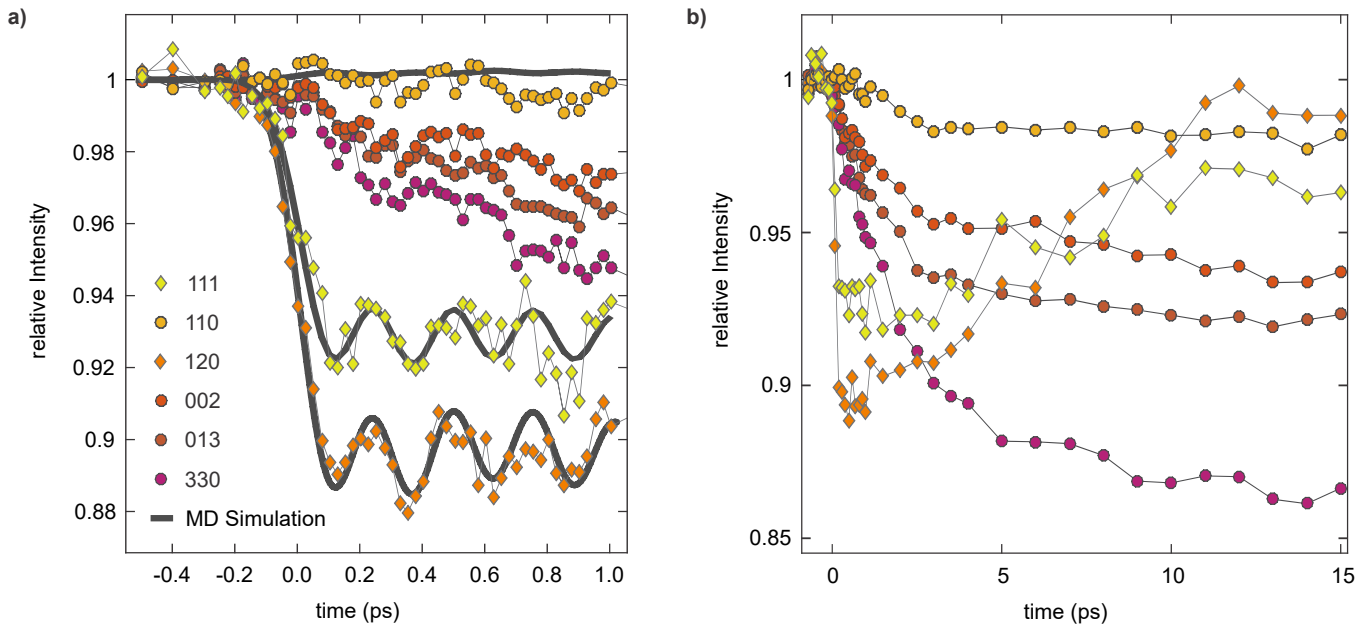}
\caption{Measured and simulated evolution of the intensity of different Bragg peaks in {\textbf a)} the first picosecond after excitation  and  {\textbf b)} in a window of 15 picoseconds. Diamond symbols are used for PD peaks and circles denote Bragg-peaks which are also present in the high-symmetry phase. The thick solid lines in a) are obtained from the MD simulations by convolution with a Gaussian function to account for the temporal resolution of the experiment. Peak labels only denote the most prominent peaks of the fit.}
\label{fig:pdecay}
\end{figure*}
%%%%%%%%%%%%%%%%%%%%%%%%%%%%%%%%%%%%%%%%%%%%%%%%%%%%%%%%%%%

The measured temporal evolution of the intensity of several Bragg-peaks after excitation with a fluence of 2.6~mJ/cm$^2$ is shown in \fig{fig:pdecay}. Two distinct processes are observed in the evolution of the peaks. Within the temporal resolution of the experiment, the PD peaks show a sudden and large decrease of intensity, which is followed by an oscillation. The intensity of the HS peaks does not abruptly change, but decreases on much longer time delays of several picoseconds on which also the intensity of the PD peaks continue to evolve. 
The immediate changes and the oscillation of the intensity of the peaks are related to the coherent real-space motion of the atoms, as this modulates the structure factor and thus the intensity of the Bragg-peaks (Eq. \ref{eq:sf}), whereas the slow decrease of intensity can be related to incoherent energy transfer to a broad range of phonons which increases the atomic mean square displacement (MSD). It is worth noting that the HS peaks of antimony do not show a fast initial drop indicative of global laser-induced phonon softening, in contrast to the ultrafast response of bismuth~\cite{Johnson2009}. 

To describe the complex lattice dynamics of antimony, including both coherently and incoherently excited phonons, Giret et al.~proposed a modified two-temperature model (TTM), in which the real-space motion of the \Ag\ mode is calculated separately from the coupled differential equations describing the incoherent energy transfer between electrons and phonons \cite{Giret2011}.
We adopt this method and calculate the changes of Bragg peak intensity, caused by the excitation of the  \Ag\ phonon mode, from our MD simulations and apply multi-temperature models to calculate the electron-phonon coupling strength from our diffraction data. 

Our MD simulations reveal that the excitation of electrons to $T_e = 2000$ K shifts the Peierls parameter from $z=0.2344$ to $z=0.2352$, corresponding to a shift of the minimum of the potential energy surface of $0.9$ pm along the z-direction towards the more symmetric phase. 
This results in a collective oscillation of atoms around this new equilibrium position. The change in diffracted intensity in three exemplary peaks ((111), (110) and (120)), calculated from the MD data as described above, is shown as solid lines in \fig{fig:pdecay}a. 
For comparison to the data, the results have been convoluted with a Gaussian function of 200 fs (FWHM) and scaled by a factor 0.9, 0.9 and 1.35, respectively. The scaling is needed because the excitation conditions in the MD simulation and the experiments are only similar to a certain extent. Additionally, the determination of an absolute change in intensity of single peaks is, despite the elaborate data analysis, difficult due to the diffracted intensity from the amorphous \SiN, which is why different scaling factors are applied. The convolution reduces the amplitude of the oscillations and is an upper limit of the time resolution of the experiment, as spatially inhomogeneous excitation \cite{KST2015} might additionally decrease the amplitude of the observed oscillations.
The simulation reveals that only the peaks related to the Peierls-like distortion show a large decrease in amplitude, whereas the HS peaks are predicted to minutely increase in intensity. The frequency at the excitation density of the experiments is extracted from the simulations to be $3.9$ THz, corresponding to a significant softening of the mode compared to the $4.5$ THz observed in experiments with low excitation densities \cite{Zeiger1992}, and in agreement with the experimental data (\fig{fig:pdecay}). The amplitude of the $E_g$ mode, which can be observed in optical and x-ray diffraction experiments \cite{Hase1996, Johnson2013}, is too small to be detected here.

We quantify the incoherent energy transfer from electrons to the lattice by solving the coupled differential equations of a TTM, which describe the changes in electronic and lattice temperature as a function of the temperature difference between the two subsystems and compute the electron-phonon coupling from our DFT simulations. The highly excited \Ag\ mode is excluded from this treatment, following Arnaud et al. \cite{Arnaud2013}. The energy content of the coherently excited \Ag\ mode is estimated to be insignificant in the entire energy-balance, justifying this approach. 
However, the assumption of a thermal phonon distribution is questionable, as the coupling of electrons to different phonon branches can vary significantly \cite{Waldecker2016, Arnaud2013}. However, as no first-principle coupling parameters have been reported to date, the TTM is the only model applicable to fit the data. 

\begin{figure*}[tbhp]
\centering
\includegraphics[width=1.0\textwidth]{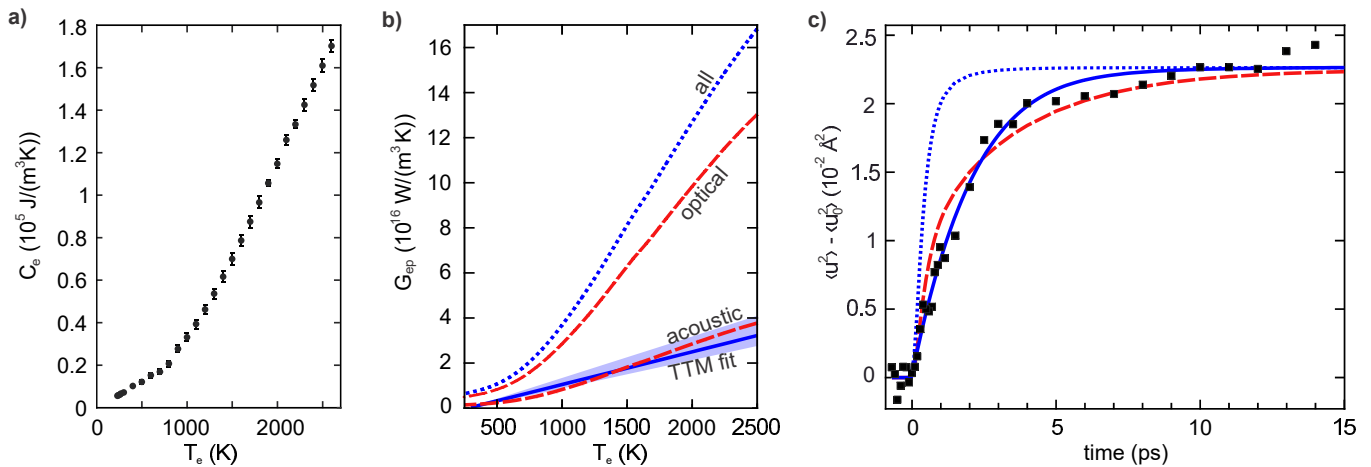}
\caption{{\textbf a)} Electronic heat capacity of Sb as a function of electronic temperature, calculated by DFT. {\textbf b)} Electron temperature dependence of the electron-phonon coupling strength. The blue solid line is the retrieved coupling strength from the TTM fit with the light blue denoting its error. The blue dashed line is $G_{ep}$ from the  the DFT calculations and the red dashed lines are the partial couplings $G_{ep,i}$ to optical and acoustic phonon branches, respectively. {\textbf c)} Measured change in atomic mean square displacement (black squares) as a function of pump-probe delay. The blue solid line is derived from the two-temperature model fit, the dotted blue line is the MSD obtained by running a TTM with the DFT values of $G_{ep}$ as an input. The red dashed line is the result of the non-thermal lattice model using the {\it ab initio} partial coupling strengths.}
\label{fig:TTM}
\end{figure*}

The calculated electron-phonon coupling (see Sec. \ref{comp-meths}) is plotted in \fig{fig:TTM} b. We find a pronounced dependence of the electron-phonon coupling on electronic temperature. 
A similarly strong dependence has been reported for bismuth \cite{Gamaly2013, Arnaud2013}, which has been explained as a consequence of the strongly varying electronic density of states near the Fermi level. 
The red dashed curves in \fig{fig:TTM} b show the partial coupling strength of electrons to the acoustic as well as to the optical phonon branches. 
For all considered electron temperatures, the incoherent electron-lattice interaction is dominated by the coupling to optical phonons. This implies that the two-temperature approximation fails in out-of-equilibrium states and that a TTM might not accurately describe the microscopic energy flow between electrons and lattice.
As the TTM is the commonly applied model, however, at first we continue by fitting a TTM to our data, compare the obtained electron-phonon coupling strength to our first principle values and discuss the implications of non-thermal phonons thereafter.

We calculate the change of atomic mean square displacement from the change of intensity of the HS Bragg peaks via the relation
\beq
\label{eq:BTt}
\langle u^2 \rangle (t) - \langle u_0^2 \rangle = - \frac{3}{4\pi^2} \frac{\ln (I_{\mathrm{rel},s}(t))}{s^2}.
\eeq
Using tabulated values for the temperature dependence of the Debye-Waller factor \cite{book:Peng}, the mean-square displacement is converted into a lattice temperature, which explicitly assumes the phonons to be in thermal equilibrium. The MSD is plotted as a function of delay time in \fig{fig:TTM} c.

The coupled differential equations of the TTM are solved and optimized numerically to best reproduce the experimental values of the lattice temperature and are given by 

\begin{subequations} \label{eq:TTM}
\begin{align}
C_e(T_e) \frac{\partial T_e}{\partial t} &= - G_{ep}(T_e) \cdot (T_e-T_l) + f(t-t_0)\,, \\
C_l(T_l) \frac{\partial T_l}{\partial t} &= G_{ep}(T_e) \cdot (T_e-T_l)\,.
\end{align}
\end{subequations}

The function $f$ models the temperature increase of the electrons by a Gaussian function, the integral of which is the absorbed energy density $u_E$. The variable parameters of the optimization are the absorbed energy density, the time of pump-probe overlap $t_0$ and the electron-phonon coupling $G_{ep}(T_e)$, see also \cite{Waldecker2016}. The electronic heat capacity $C_e(T_e)$ is calculated from our first principle DFT data via $C_e = (dE/dT_e)\vert _V$, (see Sec.\ \ref{comp-meths},) and is plotted as a function of electronic temperature in \fig{fig:TTM}~a. \added{Linear fits to the electronic heat capacity, $C_e = m \cdot T_e + b$, result in $m = (25.7 \pm 0.6)$ J/(m$^3$K$^2$) and $b = (-37 \pm 30)$ J/(m$^3$K) in the low-temperature region (below 800~K) and $m = (88.7 \pm 0.8)$ (J/m$^3$K$^2$) and $b = (-6200 \pm 200)$ J/(m$^3$K) for temperatures between 1100~K and 2600~K.} The lattice heat capacity $C_l$ was taken to be independent of temperature at a value of $C_l = 1.38\cdot10^6$ J/(m$^3$K) \cite{Wagman1982}. We include a dependence of $G_{ep}$ on $T_e$ in the numerical optimization process, which we restrict to being linear, $G_{ep}(T_e) = G_{ep,0} + g\cdot T_e$, as the data quality and the numerical optimization do not allow to fit a more complex dependence. Heat flow into the capping layers is not expected to significantly influence the evolution of the experimentally obtained lattice temperature, as it has been shown in similar experiments that the equilibration with the \SiN\ substrate occurs on delay times of several tens to hundreds of picoseconds \cite{Waldecker2015GST}, and would mostly be important in very thin films \cite{KST2015}.

%%%%%%%%%%%%%%%% Figure 3 %%%%%%%%%%%%%%%%%%%%%%%%%%%%%%%%%%%%%%%%%%%%%%%%%%%%%%%%%%%
%%%%%%%%%%%%%%%%%%%%%%%%%%%%%%%%%%%%%%%%%%%%%%%%%%%%%%%%%%%

The lattice temperature, calculated by the TTM with the optimized parameters,  reproduces well our data and is plotted as blue solid line in figure~\fig{fig:TTM}~c. 
The retrieved $G_{ep}(T_e)$ is plotted in \fig{fig:TTM}~b. The light blue area is the error of the fit, which we obtain by running the TTM with different start parameters and taking the variation of the optimized parameters. The experimentally retrieved $G_{ep}(T_e)$ is significantly smaller than the one calculated by DFT. 
Similarly, the evolution of the MSD calculated by a TTM with the DFT results for $G_{ep}(T_e)$ does not reproduce our data (see dotted line in \fig{fig:TTM} c). 
We explain this quantitative difference between the theoretically and experimentally obtained values of $G_{ep}(T_e)$ as failure of the two-temperature approximation as a result of the markedly different coupling to optical and acoustic phonons (\fig{fig:TTM} b). After optical excitation of the electrons, incoherent electron-phonon coupling preferentially generates optical phonons. High-energy phonons, however, contribute less to the atomic MSD compared to low-energy phonons~\cite{Waldecker2016}. In such a case, the analysis of experimentally determined MSD with the assumption of thermal phonon distributions results in an underestimation of the energy content of the phonon degrees of freedom, and consequently in an underestimation of the electron-phonon coupling strength. 

We therefore apply a refined model lifting the restriction to thermal phonon distributions (non-thermal lattice model, NLM) by splitting the phonons into subgroups~\cite{Waldecker2016}, and assigning temperatures to these smaller subgroups of phonons. 
Here, we treat acoustic and optical phonons independently and calculate the energy balance between electrons and both types of phonons as three independent subsystems:
\begin{subequations} \label{eq:NLM}
\begin{align}
C_e(T_e) \frac{\partial T_e}{\partial t} &= \sum_{i=a,o} - G_{ep,i}(T_e) \cdot (T_e-T_i) + f(t-t_0)\,, \\
C_i(T_i) \frac{\partial T_i}{\partial t} &= G_{ep,i}(T_e) \cdot (T_e-T_i) + G_{pp} \cdot (T_i-T_j)\,,
\end{align}
\end{subequations}
where $i,j=\{a,o\}$ and $i \neq j$. The heat capacity of the acoustic ($a$) and optical ($o$) phonon branches are calculated to be $6.42\cdot10^5$~J/m$^3$K and $6.31\cdot10^5$~J/m$^3$K, respectively. 
We solve the NLM numerically and compare to the experimental data by computing the MSD from the temperatures of both phonon subsystems~\cite{Waldecker2016}. 
All material-specific parameters are taken from first principles calculations except for the phonon-phonon coupling parameter $G_{pp}$ connecting the two phonon systems, which is optimized numerically. 
%From the temperatures of the phonon branches, their individual contributions to the MSD is calculated. 
The best fit is obtained for $G_{pp} = 8\pm1\cdot10^{16}$ W/m$^3$K, see red dashed line in \fig{fig:TTM} c. 
The NLM employing the first principle values of $G_{ep,i}$ reproduces the data much better than the TTM with the calculated $G_{ep}$, as shown in \fig{fig:TTM} c. 
This suggests that the minimal extension of the TTM by separate treatment of acoustic and optical phonons results in a significantly more accurate description of the energy flow between electrons and lattice. 
The NLM is a minimal extension of the TTM and does not account for non-thermal electronic distribution or arbitrary non-thermal phonon distributions \cite{Chase2016}, which likely explains the remaining deviation of data and NLM at short times. Our observations imply, however, that the NLM is a sufficient extension of the NLM for a quantitative description of the energy exchange between electrons and lattice.

\section{Summary and Conclusions}

We have investigated the lattice dynamics of antimony in laser-induced nonequilibrium states. By applying femtosecond electron diffraction experiments with tilted, polycrystalline but textured samples, and by performing ab initio molecular dynamics simulations, we have determined the effects of coherent as well as incoherent electron-phonon coupling in a single measurement. 
For a fluence of 2.6~mJ/cm$^2$, our MD simulations predict that the laser-excitation leads to a shift of the potential-energy surface of $0.9$ pm, resulting in a coherent atomic oscillation of the \Ag\ mode of 3.9 THz. 
The spatial and temporal resolution of the compact FED experiment are sufficient to resolve this motion, which is the first observation of the oscillations caused by a fully symmetric coherent optical phonon in a solid with an electron diffraction experiment. 
In addition, the incoherent energy-transfer of electrons to the lattice is analyzed in the framework of a TTM and a model allowing for non-thermal phonon distributions. 
The evolution of the lattice temperature is extracted from the data, and the electron-phonon coupling parameter $G_{ep}$ is determined from the measurements as well as from {\it ab initio} DFT calculations. 
Using the TTM with $G_{ep}$ as free fit parameter results in a very good agreement with the measured time-dependent MSD. In comparison to DFT calculations, however, the TTM analysis significantly underestimates the electron-phonon coupling strength. We attribute this to the large differences in coupling strength of electrons to acoustic and optical phonon branches and a failure of the two-temperature approximation.
By allowing for non-equilibrium between these phonon branches in a non-thermal lattice model, the DFT calculations are able to reproduce the data.
This work demonstrates the suitability of FED measurements combined with ab initio simulations to probe coherent and incoherent microscopic coupling mechanisms and elementary atomic motion in solid state systems exhibiting complex dynamics.
\\

\section{Acknowledgments}

We thank Dhriti Ghosh and Valerio Pruneri (ICFO Castelldefels) for supplying the samples. The work was financed by the Max Planck Society and by the DFG projects GA 465/16-1 and ZI 1307/1-1. Computations were performed at the Lichtenberg High Performance Computer of the Technical University Darmstadt (Project ID 130). R. B. thanks Alexander von Humboldt foundation for financial support. 

%\bibliography{Sb_literature}

%

\end{document}